\documentclass[%
 reprint,
 amsmath,amssymb,
 aps,
]{revtex4-2}

\usepackage{xcolor}
\usepackage{graphicx}
\usepackage{dcolumn}
\usepackage{bm}
\usepackage{soul} 
\usepackage{siunitx}
\usepackage{braket}
\usepackage{comment}

\begin{document}

\preprint{APS/123-QED}

\title{High-Q superconducting microwave resonators using MBE titanium nitride}

\author{Anand Ithepalli$^{1,}$}
\thanks{Equal contribution}
\author{Haoran Lu$^{2,}$}
\thanks{Equal contribution}
\author{Eegene Clara Chung$^{3}$}
\thanks{Equal contribution}
\author{Xiangqin Wang$^{2}$}
\author{Amit Rohan Rajapurohita$^{2}$}
\author{Keun-Yeol Park$^{4}$}
\author{Celesta S. Chang$^{4}$}
\author{Peter McMahon$^{2,5}$}
\author{Huili Grace Xing$^{1,5,6}$}
\author{David Muller$^{2,5}$}
\email{david.a.muller@cornell.edu}
\author{Valla Fatemi$^{2}$}
\email{vf82@cornell.edu}
\author{Debdeep Jena$^{1,5,6,}$}
\email{djena@cornell.edu}

\affiliation{$^1$Department of Materials Science and Engineering, Cornell University, Ithaca, NY 14853, USA\\$^2$School of Applied and Engineering Physics, Cornell University, Ithaca, NY 14853, USA\\$^3$Department of Physics, Cornell University, Ithaca, NY 14853, USA\\$^4$Institute of Applied Physics, Department of Physics and Astronomy, Seoul National University, Seoul 08826, South Korea\\$^5$Kavli Institute at Cornell for Nanoscale Science, Cornell University, Ithaca, NY 14853, USA\\$^6$School of Electrical and Computer Engineering, Cornell University, Ithaca, NY 14853, USA}


\begin{abstract}
Using molecular beam epitaxy, we have realized thin films of titanium nitride (TiN) on c-plane sapphire that exhibit the lowest observed full-width at half maximum X-ray rocking curve width of 18 arcsec.  Though the (111) oriented TiN exhibits an abrupt and crystalline interface with sapphire, for the first time we observe sub-surface defects in the sapphire substrate, which nucleate structural defects in the epitaxial TiN layer.  Using quarter-wavelength coplanar waveguide (CPW) resonators in a 3 \textmu m/6 \textmu m/3 \textmu m gap/strip/gap lines in a hanger geometry, we find the internal quality factor of the TiN resonators to be $>10^{6}$ in the single-photon $\langle n \rangle \sim 1$ limit at 5.8 GHz and 10 mK, rising to $>20 \times 10^{6}$ at $\langle n \rangle \sim 10^{6}$.  The results are of high interest for applications of superconducting TiN in several areas, and provide a path towards epitaxial Josephson junctions with crystalline barriers in the future for high coherence qubits. 


\end{abstract}

\maketitle

{\bf {\em Introduction:}} The remarkable electronic, thermal, mechanical, chemical, refractory, and plasmonic properties of TiN thin films drive its heavy usage in silicon CMOS processing of electronic and photonic integrated circuits \cite{boltasseva2015all}.  Because TiN thin films exhibit superconductivity at cryogenic temperatures, its practical value makes it of high interest for a wide range of superconducting devices and circuits \cite{vissers2010low, gao2012titanium}.  There is a rejuvenated interest in TiN \cite{gao2022epitaxial, tominaga2025intrinsic, smart2026thermal, terai2026superconducting} for superconducting device applications. 
This is due to the possibility of all-epitaxial superconducting microwave qubits using TiN and MgO which possess the same crystal symmetry and lattice mismatch less than $0.5\%$ \cite{takiguchi2025electronic}. Such a structure can potentially eliminate losses and decoherence that originate from amorphous tunnel barriers needed in qubits \cite{martinis2005decoherence, oh2006elimination}.

Towards this goal, here we show that MBE-grown TiN films on c-plane sapphire exhibit state-of-the-art material properties and microwave performance.  A sharp superconducting transition with $\Delta T_{\text{c}}<0.03$ K is observed around $T_{\text{c}} \simeq 5.1$ K. Coplanar waveguide (CPW) superconducting microwave resonators are widely used as a proxy to assess the quality factor of base layers for superconducting qubits \cite{mcrae2020materials, gao2022epitaxial, crowley2023disentangling, olszewski2026low, mcfadden2025interface}. CPW resonators fabricated from the TiN reveal averaged internal quality factor $Q_{\rm i}>10^6$ at 10 mK in the single-photon limit {\em without} any intentional surface oxide removal by acid-treatment nor any overetch into the substrate.  We first describe the epitaxial growth, structural and electronic properties of TiN, defects in the epitaxial structures, and microwave characteristics of resonators made of them.  Then we compare our results with those in the literature. We conclude by discussing the path towards all-epitaxial crystalline qubits.

\begin{figure*}[t!]
    \centering
    \includegraphics[width=0.9\textwidth]{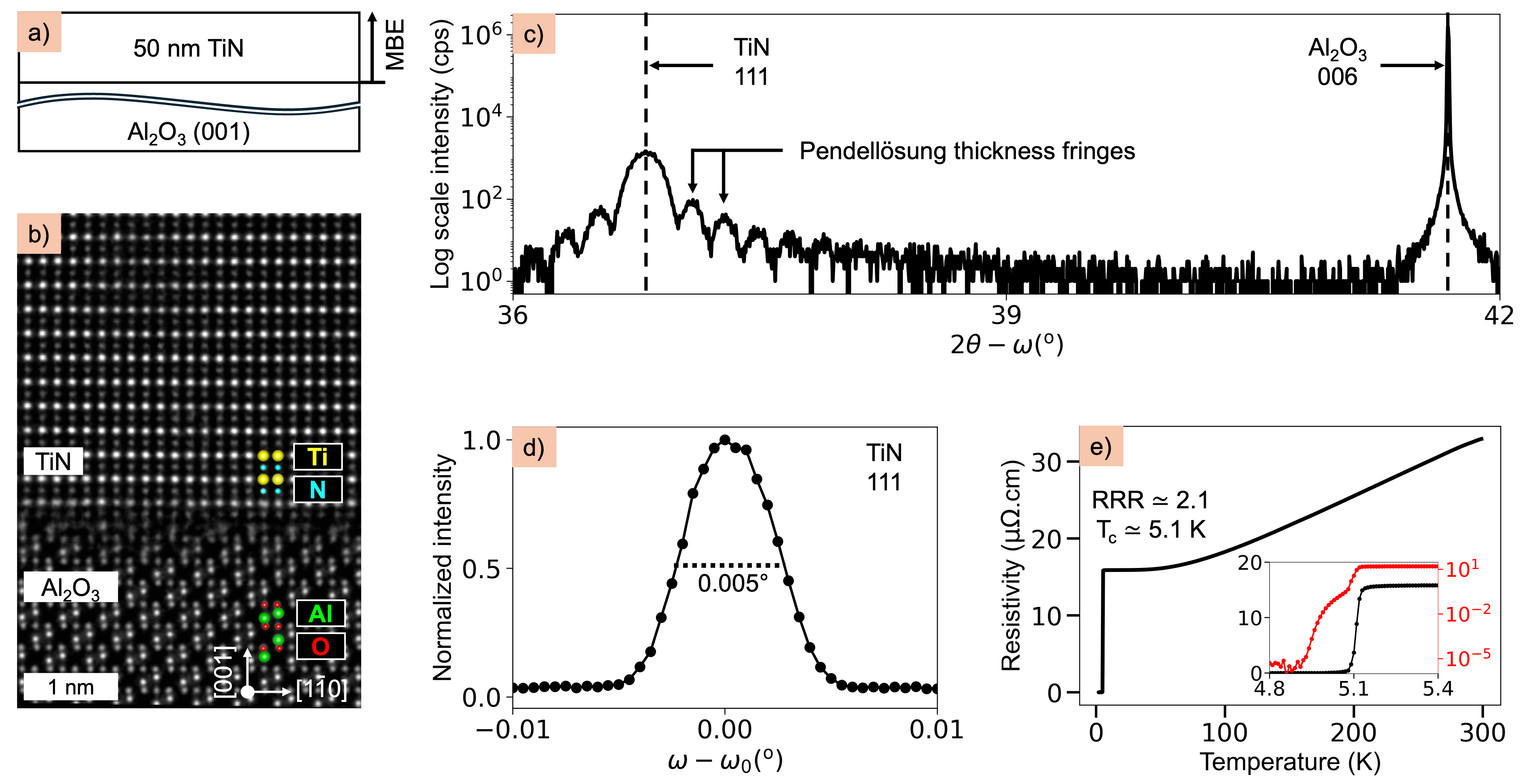}
    \caption{(a) Sample schematic of the MBE-grown TiN film used in this study. (b) Experimental multislice electron ptychography (MEP) STEM image of the film shown in a) with the lattice models of both TiN and sapphire. (c) Symmetric coupled 2$\theta-\omega$ XRD scans indicating a single-crystal TiN film. (d) Rocking curve around the symmetric TiN 111 reflection with a record-low FWHM of 18 arcsecs. e) Temperature-dependent resistivity of the 50 nm TiN film indicating a residual resistivity ratio of 2.1 with inset (e) showing the superconducting transition around 5.1 K.}
    \label{fig:MatProps}
\end{figure*}
{\bf {\em MBE growth of TiN:}} The TiN discussed in this work is grown using MBE on (001) (Al$_2$O$_3$) substrates. 
Ganjam \textit{et al.} observed that annealing of sapphire reduces its cryogenic microwave loss tangent from $\tan(\delta) \sim 2 \times 10^{-7} \rightarrow 2 \times 10^{-8}$ \cite{ganjam2024surpassing}. Therefore, we used commercially annealed sapphire as the starting substrate. Prior to TiN growth, the substrate is cleaned for 5 minutes in acetone followed by 5 minutes in 2-propanol in an ultrasonic bath to dissolve both polar and nonpolar contaminants. The wafer is then rinsed in deionized (DI) water and blow-dried using N$_2$ gas. The dried wafer is cleaned in piranha solution (2:1 ratio of H$_2$SO$_4$:H$_2$O$_2$) for 20 minutes at an average temperature of 90$^\circ$C to remove all organic contaminants similar to \cite{ganjam2024surpassing}. The piranha-cleaned wafer is then rinsed in two beakers of fresh DI water and blow-dried using N$_2$ gas again.

The cleaned substrate is transferred to the growth chamber of a Veeco GenXplor MBE system through a loadlock (LL) and a buffer chamber. After loading the substrate, the LL chamber is baked at 200$^\circ$C for 7 hours and cooled down to room temperature. The base chamber pressures of the LL, buffer and the growth chambers during the substrate transfer are $\sim 1\times10^{-8}$ Torr, $\sim 1\times10^{-10}$ Torr, and $\sim 1\times10^{-10}$ Torr respectively.


The substrate is heated to 600$^\circ$C in the growth chamber, as measured by a thermocouple. 6N5 pure N$_2$ gas was introduced into the chamber at a flow rate of 2 standard cubic centimeters per minute (sccm). Active nitrogen is generated using radio-frequency generator operated at 200 W. 4N5 pure Ti source is evaporated using a Telemark 4-pocket ebeam evaporator connected to the MBE system. Ti flux is monitored during the growth using electron impact energy spectroscopy (EIES) and kept at a constant value to obtain 2.5 nm/min growth rate of TiN. Active nitrogen to Ti flux ratio is around 1.5 during the film growth. With these conditions, the TiN film is grown for 20 minutes to obtain a 50 nm layer as shown in Fig.~\ref{fig:MatProps}(a).



{\bf {\em Structural and electronic properties of the TiN film:}} Fig.~\ref{fig:MatProps}(b) shows the experimental image of TiN-sapphire interface using multislice electron ptychography (MEP). MEP is a scanning transmission electron microscopy (STEM)-based imaging technique that reconstructs the atomic potentials in three dimensions (3D) with sub-Ångstrom lateral resolution and 2-4 nm depth resolution \cite{chen2021electron}. The rocksalt structure of Ti and N atoms, as well as the corundum structure of Al and O atoms in the sapphire, are clearly resolved. An atomically sharp interface is observed between TiN and sapphire. This is further supported by Pendell\"osung fringes observed in coupled symmetric 2$\theta-\omega$ x-ray diffraction (XRD) scan collected using a table-top Malvern Panalytical high-resolution XRD tool. The XRD scan is shown in Fig.~\ref{fig:MatProps}(c). The rocking curve (RC) in Fig.~\ref{fig:MatProps}(d), collected around the symmetric 111 reflection of TiN, indicates a full width at half maximum (FWHM) of 0.005$^\circ$ or 18 arcsecs. This is the lowest ever reported value of RC FWHM for TiN thin films, 8 times lower than the previous best value of 0.04$^\circ$ for 100 nm TiN sputtered on c-plane sapphire \cite{gao2022epitaxial}. 
This indicates that TiN(111) planes are extremely parallel to the (001) planes of sapphire. Atomic force microscopy showed 0.5 nm surface roughness of the TiN layer, and river-rock morphology similar to MBE TiN on Si \cite{richardson2020low}.  The effect of the RC and surface roughness on the microwave loss of TiN is not yet established.

\begin{figure}[t!]
    \centering
    \includegraphics[width=0.4\textwidth]{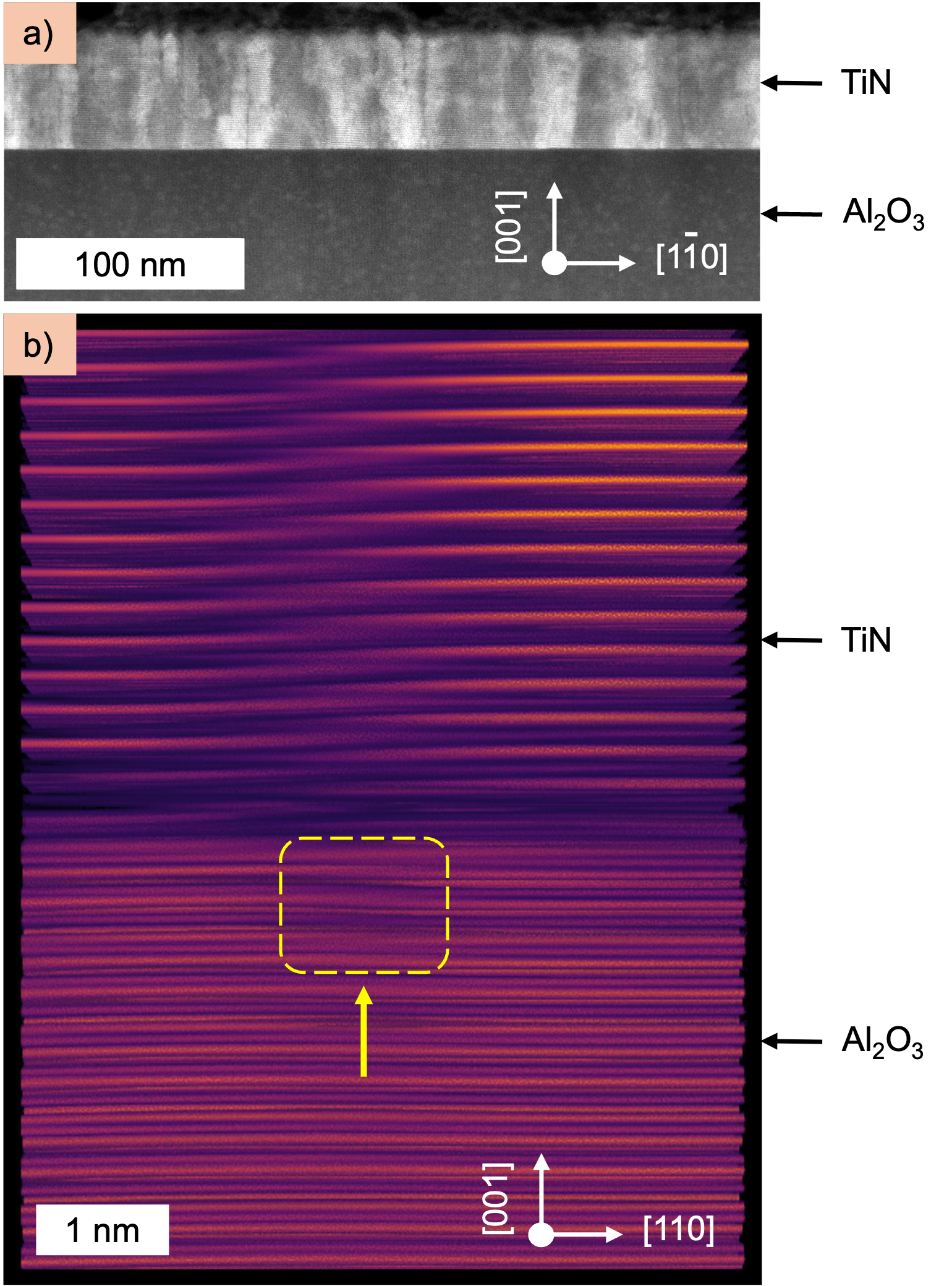}
    \caption{(a) Columnar growth of TiN is seen from a broader view of the HAADF-STEM image. (b) 3D visualization of MEP reconstruction showing a different zone axis compared to that of a). The yellow box and arrow indicate the subsurface dislocation in sapphire which nucleates a screw-type dislocation in TiN film.}
    \label{fig:STEM}
\end{figure}

The temperature dependent electrical resistivity shown in Fig.~\ref{fig:MatProps}(e) is measured from a Hall-bar with 40 squares between the voltage probes. Room temperature resistivity $\rho_\text{300 K}=$ 33 \textmu$\Omega$.cm and the residual resistivity ratio (RRR) $\rho_\text{300 K}/\rho_\text{10 K}=$ 2.1 are measured from this TiN film. These are typical values for MBE-grown TiN films at a relatively low growth temperature of 600$^\circ$C on c-plane sapphire \cite{olson2015growth}. The inset of Fig.~\ref{fig:MatProps}(e) shows a narrow superconducting transition around 5.1 K in linear scale (black). The logarithmic y-axis of the same inset shows a broadened tail up to 4.9 K. Such broadness is typically attributed to disorder in superconducting films as opposed to nontrivial physics around critical points of phase transitions \cite{StrunkBKT}.


{\bf {\em Advanced structural analysis of the TiN film:}} Fig.~\ref{fig:STEM}(a) shows the high-angle annular dark-field (HAADF) STEM image of the TiN film on sapphire over a larger field of view. The growth of a rocksalt symmetry TiN on hexagonal symmetry sapphire promotes formation of twin domains, which is observed during MBE growth using \textit{in-situ} reflection high-energy electron diffraction similar to earlier works of TiN \cite{olson2015growth, richardson2020low}. Columnar grains are seen in the TiN film due to low temperature and nitrogen rich growth conditions required to reduce N-vacancies \cite{olson2015growth, takiguchi2025electronic}. Despite twinning and columnar features, the RC FWHM around the symmetric peak of TiN films is consistently shown to be lower on c-plane sapphire than the lattice and symmetry matched MgO substrates \cite{khim2023electrical}.

MEP reconstruction of the interface is projected along a [110] zone-axis in Fig.~\ref{fig:STEM}(b) which is 90$^\circ$ away from the [1$\overline{1}$0] zone axis of Fig.~\ref{fig:STEM}(a). The image shows an edge-type dislocation with a Burgers vector parallel to the growth direction in the subsurface of the substrate indicated in the dashed box. Interestingly, a screw-type dislocation in the TiN is also observed to have nucleated at the same position with the Burgers vector parallel to the growth direction. While the exact origin of the subsurface dislocation in sapphire substrate is unknown, it is rather common for the commercially polished substrates to have damage extending few nanometers below surface \cite{pinkas2010thermal}. Annealing the substrates post-polishing reduces this damage through recrystallization process \cite{pinkas2010thermal}. Potential impact of such dislocations is discussed in the supplementary material. Note that while the lateral sampling of the MEP reconstruction is 0.02 nm as in Fig.~\ref{fig:MatProps}(b), the depth sampling is 0.85 nm. Due to information transfer limits inherent to the technique, a finer sampling will not improve the resolution. Fig.~\ref{fig:STEM}(b) shows the depth view interpolated to match the lateral scale, hence the lack of atomic resolution in this view.




\begin{figure}[t!]
    \centering
    \includegraphics[width=0.5\textwidth]{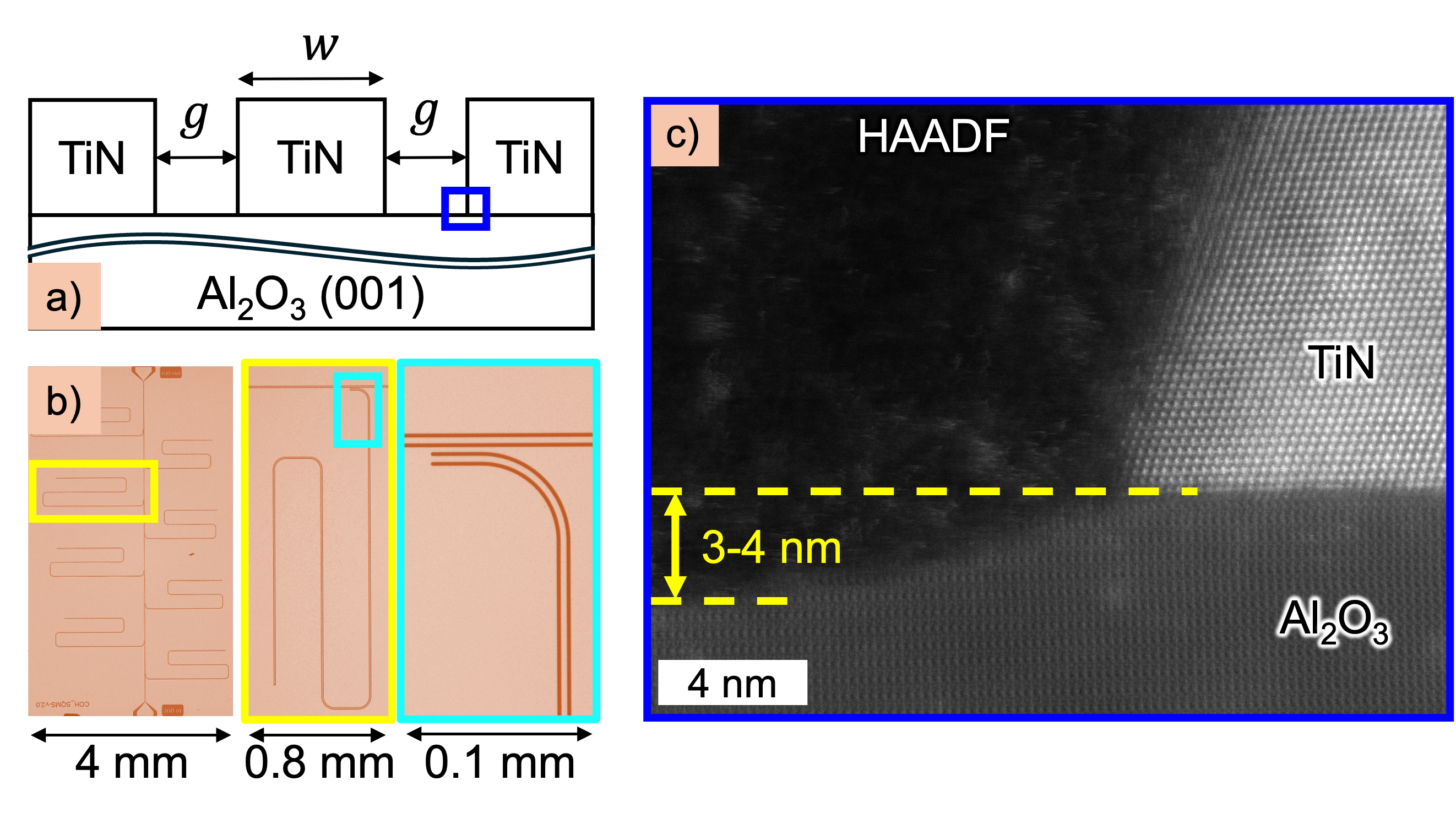}
    \caption{(a) Cross-section schematic of the processed resonator from the 50 nm TiN film. The central signal line is $w=6$ \textmu m wide and the gaps between the signal line and ground planes are $g=3$ \textmu m. (b) Representative optical images at various magnifications of a processed resonator chip with the same mask used for the sample discussed in this work. (c) HAADF image of the same sidewall region of the processed chip, viewed along sapphire [110] axis. Although sapphire acts as a natural etch stop, 3-4 nm of sapphire seems to be etched in the trenches, which has a negligible effect on the metal-air and substrate-air loss energy participation ratio, compared with no overetch case, per HFSS simulation.
    }
    \label{fig:ResonatorFab}
\end{figure}

{\bf {\em Resonator fabrication:}}  To evaluate the microwave performance of these MBE-grown TiN superconducting thin films, we fabricated two chips shown in Fig.~\ref{fig:ResonatorFab} following the open-source resonator, radio-frequency (RF) package design from the Boulder Cryogenic Quantum Testbed \cite{kopas2022simple}. 
Both chips in this test underwent the same photolithography and etching procedures, but different photoresist removal (stripping) processes. Detailed fabrication process can be found in the supplementary material. The resonator chips will be referred to as `1165 chip' and `AZ300T chip', named after the stripping solutions used for the chips. 
\begin{figure*}[t!]
    \centering
    \includegraphics[width=0.95\textwidth]{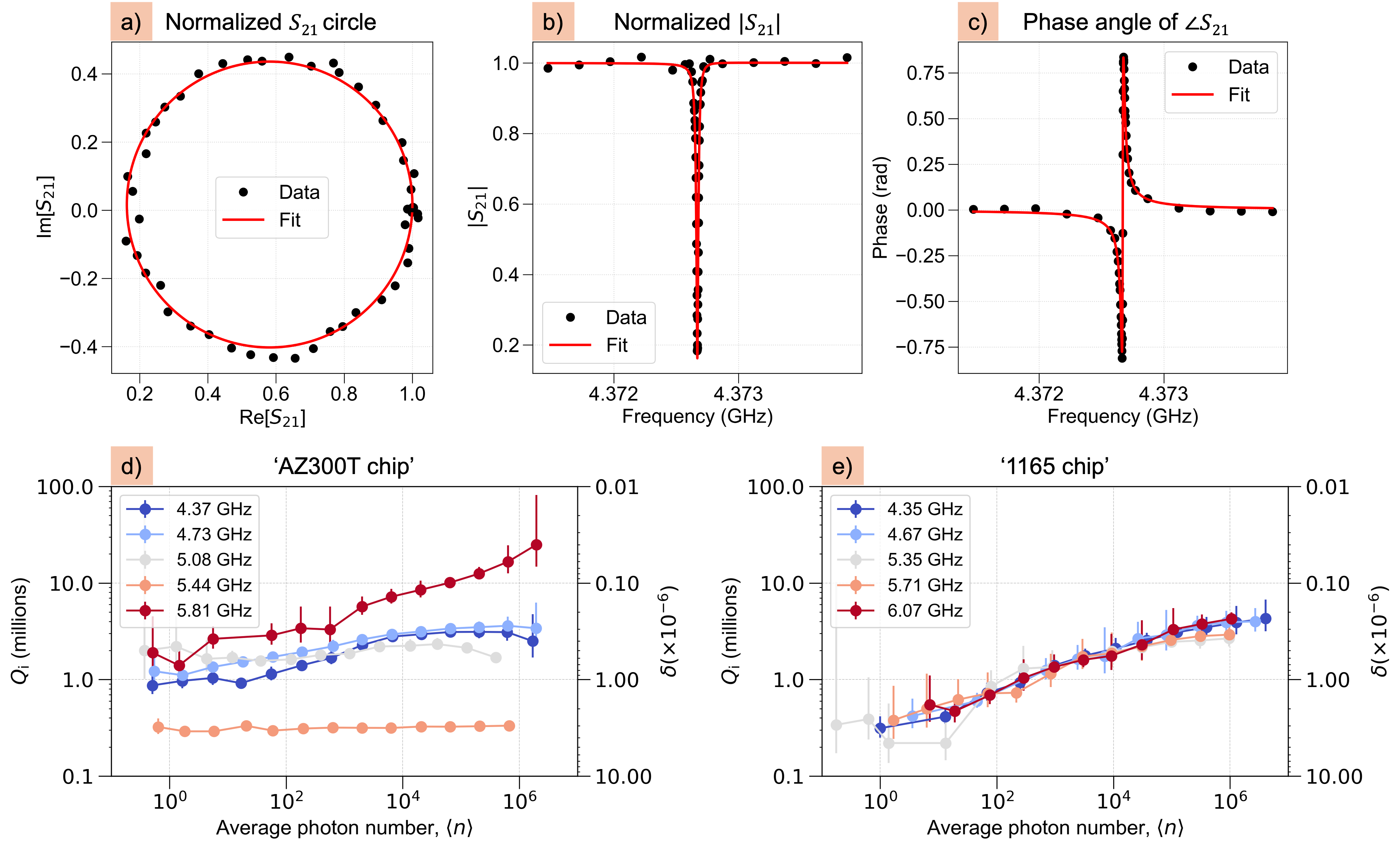}
    \caption{Measured data (black, solid circles) of the TiN resonator with $f_0 \simeq 4.37$ GHz that underwent AZ300T photoresist stripping bath in the single-photon regime, and fitting curves (red, solid lines) of: (a) normalized $S_{21}$ circle in the Argand plane, (b) normalized magnitude $|S_{21}|$ frequency dependence, and (c) phase angle $\angle S_{21}$ frequency dependence. Internal quality factor vs average photon number of TiN resonators from (d) the `AZ300T chip' and e) the `1165 chip'.}
    \label{fig:ResonatorMeas}
\end{figure*}

{\bf {\em Resonator measurement:}}  The chips are wire-bonded to the RF package using Al wires following the wire-bonding scheme in \cite{kopas2022simple}. The microwave characterization is conducted in a Bluefors dilution fridge with base temperature 10 mK. The fridge wiring is identical to that labeled as `\textbf{Fatemi}' fridge in \cite{olszewski2026low}. 
On the input side, a total of \SI{60}{\deci\bel} attenuation from multiple attenuators at different temperature flanges, one \SI{12}{\giga\hertz} KL low-pass filter, and one Eccosorb filter are placed. On the output side, one Eccosorb filter, one \SI{12}{\giga\hertz} KL low-pass filter, and one \SI{4}{\giga\hertz} to \SI{12}{\giga\hertz} isolator are placed. A copper mountain M5180 vector network analyzer with dynamic range \SI{0}{\deci\bel m} to \SI{-50}{\deci\bel m} and a programmable attenuator with tunable attenuation range \SI{0}{\deci\bel} to \SI{65}{\deci\bel} are involved for resonance peak characterization from low-photon ($\braket{n}<10$) to high-photon ($\braket{n}>10^5$) range. 
The transmission coefficient near resonance, defined using the diameter correction method (DCM), is:

\begin{equation}
    S_{21} (f) = 1-\frac{\frac{Q}{Q_{\rm c}{\rm e}^{-i\phi}}}{1+2iQ\frac{f-f_0}{f_0}},
    \label{eq-DCM}
\end{equation}

\begin{equation}
    \frac{1}{Q} = \frac{1}{Q_{\rm i}} + \frac{1}{Q_{\rm c}},
    \label{eq-Qi}
\end{equation}
where $f_0$ is central frequency, $Q_{\rm c}$ is coupling quality factor, $Q$ is loaded quality factor, and $\phi$ is the DCM phase \cite{khalil2012analysis}. 
These parameters are extracted from fitting the measured transmission coefficient $S_{21}$ as a function of input signal frequency $f$.
Internal quality factor $Q_{\rm i}$ is then extracted from Eq. \ref{eq-Qi}. The loss tangent of the resonator is defined as $\delta \simeq \tan \delta= 1/Q_{\rm i}$.

Each resonator chip consists of 8 $\lambda/4$ resonators, with gap $g=\SI{3}{\micro\meter}$ and metal pad $w=\SI{6}{\micro\meter}$, as shown in Fig.~\ref{fig:ResonatorFab}(a-b). The designed resonance frequencies are within $\SI{4}{\giga\hertz}$ to $\SI{8}{\giga\hertz}$, common frequency range for transmon qubits \cite{krantz2019quantum}. 
A witness chip that went through the AZ300T process after patterned etch is imaged using STEM to understand the effect of etch on the sidewalls. Although sapphire acts as a natural etch stop in this gas mixture, an etch into the substrate up to 4 nm is observed in the cross-sectional HAADF-STEM image shown in Fig.~\ref{fig:ResonatorFab}(c). A strict interface of TiN and sapphire even near the sidewall is also apparent from Fig.~\ref{fig:ResonatorFab}(c).



{\bf {\em Microwave characterization:}}  An example set of measured (black, solid circles) and fitted (red, solid line) $S_{21}$ of a resonator from the `AZ300T chip' with $f_0 \simeq 4.37$ GHz in the low-photon regime is shown in Fig.~\ref{fig:ResonatorMeas}. The Argand plane projection, frequency dependence of amplitude $|S_{21}|$, and phase $\angle S_{21}$ are shown respectively in Fig.~\ref{fig:ResonatorMeas}(a-c).

\begin{figure*}[t!]
    \centering
    \includegraphics[width=0.95\textwidth]{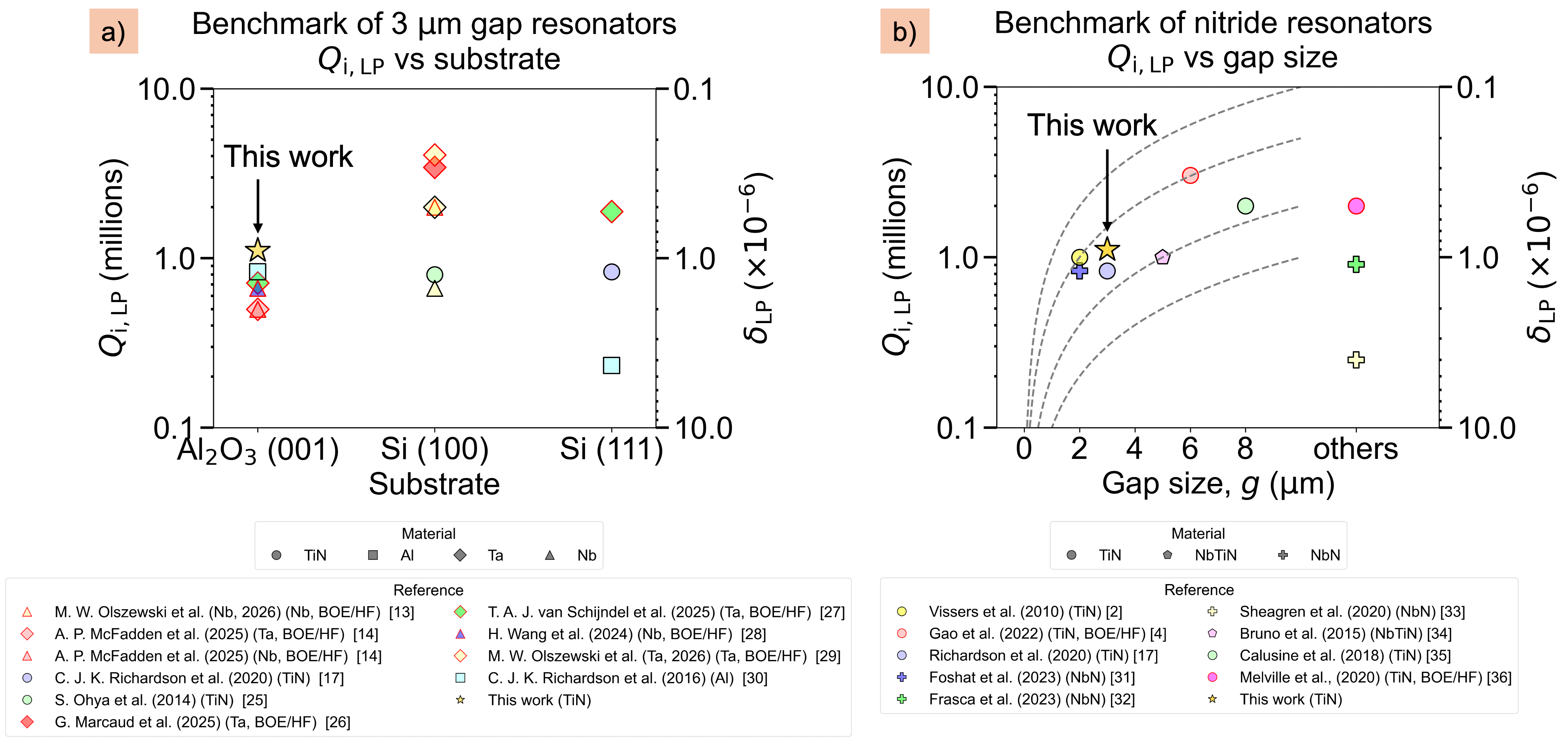}
    \caption{Benchmark plots of representative results from literature (a) low power $\langle n \rangle \simeq 1$ intrinsic loss of thin film superconductors with $g=3$ \textmu m vs substrate choices for widely used superconductors taken from references \cite{richardson2020low, ohya2013room, mcfadden2025interface, marcaud2025low, van2025cryogenic, wang2024impact, olszewski2026low, olszewski2026krypton, richardson2016fabrication} and b) low power $\langle n \rangle \simeq 1$ intrinsic loss of nitride thin film superconductors vs gap $g$ \cite{richardson2020low, foshat2023characterizing, frasca2023nbn, sheagren2020atomic, bruno2015reducing, gao2022epitaxial, calusine2018analysis, vissers2010low, melville2020comparison}. The gray dashed lines in (b) correspond to $\delta_{\text{LP}} \propto g$. The references where $g$ is not reported or is not applicable are listed on the ``others" gap size. Red outline of the markers indicates that resonators are treated with HF or BOE solutions in those studies to remove the native oxides.}
    \label{fig:Benchmarks}
\end{figure*}

$Q_{\rm i}$ vs $\braket{n}$ of the resonators on the `AZ300T chip' and the `1165 chip' are shown in Fig.~\ref{fig:ResonatorMeas}(d, e) respectively. The right y-axis of both these plots shows the loss tangent. Resonators on the `AZ300T chip' have an average $Q_{\text{i, LP}} \simeq 1.26(0.45)\times 10^6$ in the low-photon limit $\braket{n}\sim 1$ with a maximum value of $\simeq 2.20(1.12)\times 10^6$. An average $Q_{\text{i, HP}} \simeq 5.11(1.13)\times 10^6$ is observed in the high-photon limit $\braket{n}\sim 10^6$ with a maximum value of $\simeq 25.0(17.3)\times 10^6$.
Resonators on the `1165 chip' have an average $Q_{\text{i, LP}} \simeq 0.4(0.10)\times 10^6$ in the low-photon limit and $Q_{\text{i, HP}} \simeq 3.63(0.43)\times 10^6$ in the high-photon limit.


Improvement in average $Q_{\text{i, LP}}$ by switching the photoresist stripper from 1165 to AZ300T was previously reported on Nb resonators deposited on Si \cite{olszewski2026low}, and was correlated with the removal of excess Cl$^-$ ions incorporated on the sample surface after etching in the BCl$_3$ + Cl$_2$ chemistry using AZ300T solution. Across the resonators, the spread in the $Q_{\rm i}$ vs $\braket{n}$ curves is negligible in the `1165 chip' compared to the `AZ300T chip'. While the average $Q_{\text{i, LP}}$ has seen a significant improvement by a factor of 3 with the use of AZ300T solution, the standard deviation is a factor of 4 lower in the case of the `1165 chip'.

The $Q_{\text{i, LP}}$ reported in this work is among the highest values reported in the literature of CPW resonators on sapphire made from various superconductors with $g=3$ \textmu m as shown in the benchmark Fig.~\ref{fig:Benchmarks}(a). Even among nitride-based superconducting resonators with various gap widths $g$, as shown in the benchmark Fig.~\ref{fig:Benchmarks}(b), results from this work are close to the state-of-the-art values.
The red outline of markers in Fig.~\ref{fig:Benchmarks}(a, b) indicates diluted hydrofluoric acid (HF) or buffered oxide etchant (BOE) treatment of the resonator chips prior to wirebonding and measurement. Such a process is beneficial in improving the resonator quality factors due to reduced surface losses by thinning or modifying native surface oxides.
However, dilute HF and BOE are not compatible with conventional Al/AlO$_{\rm x}$/Al JJs currently necessary for qubits. 

In this work, neither of the resonator chips has been subjected to HF/BOE treatments.
It has also been shown for Ta-based resonators that performance immediately after an AZ300T strip bath is consistent with performance of resonators that witness the entire fabrication process, suggesting that our resonator performance will translate to performance of base layers for superconducting qubits~\cite{olszewski2026krypton}. 

Regardless of the surface treatments, the most popular superconductor choices are Al, Nb, Ta and TiN due to their repeatable demonstration of high $Q_{\rm i} \sim 10^6$ in the single-photon limit \cite{richardson2020low, ohya2013room, mcfadden2025interface, marcaud2025low, van2025cryogenic, wang2024impact, olszewski2026low, richardson2016fabrication}. While elemental metals like Al, Nb, and more recently, Ta have shown great promise, there is currently no approach of a crystalline barrier for a Josephson junction with these base metals. Whereas, nitride materials have been shown to have a few choices of crystalline stacks like NbTiN/AlN/NbTiN \cite{cyberey2018nbtin, supple2024atomic}, NbTiN/GaN/NbTiN \cite{cyberey2023supergan}, NbN/AlN/NbN \cite{kim2021enhanced, wang2026all}, and TiN/AlN/TiN \cite{nakayama2016fabrication}. 
NbN/MgO/NbN and TiN/MgO/TiN junctions are of utmost interest due to NbN, TiN, and MgO possessing the same rock-salt crystal structure  \cite{shoji1985niobium, terai2026superconducting}. Further, TiN and MgO have only $\simeq 0.5\%$ lattice mismatch allowing for excellent epitaxy \cite{takiguchi2025electronic}. 


{\bf {\em Conclusion and outlook:}}  We demonstrated $Q_{\text{i}}$ above a million in the single-photon regime for CPW resonators made from MBE-grown superconducting TiN film with single crystal orientation on c-plane of sapphire. A change in the fabrication protocol, that is switching to AZ300T photoresist stripper from 1165 solution, has shown a factor of three improvement in the $Q_{\text{i, LP}}$ consistent with a previous report of Nb resonators on Si \cite{olszewski2026low}. Further improvement in the TiN film properties, and by extension of $Q_{\text{i, LP}}$, is possible through growth optimization as our film has lower RRR and higher room-temperature resistivity compared to the best reported values of TiN on lattice and symmetry-matched MgO substrates. The single crystal nature of the TiN is attractive for an all-crystalline superconducting qubit stack with reduced TLS defect densities.

Despite the high quality factors achieved in this work, we expect room for improvement due to the following reasons. 
\textcolor{black}{The shallow overetch on sapphire substrate, shown in Fig.~\ref{fig:ResonatorFab}(c), results in a much higher energy participation ratio ({\em{pr}}) in the lossy dielectric regions namely metal-air-substrate tri-junction (tri) and metal-air (MA), compared with those on silicon substrate with trenching~\cite{woods_determining_2019,calusine_analysis_2018}.}
Further, the etch damage of sapphire as seen from Fig.~\ref{fig:ResonatorFab}(c) may cause amorphization and increase the SA interface loss tangent while having negligible effect on the participation ratio of the microwave energy due to low etch depth of only 3 to 4 nm. Numerical results of participation ratios from Ansys HFSS$^{\text{\texttrademark}}$ simulations are presented in the supplementary material.

TiN on sapphire has high $Q_{\text{i}}$ despite the columnar structure. The 2D projection of the HAADF-STEM image in Fig.~\ref{fig:STEM}(a) enhances the contrast of the columnar grain boundaries. A large field-of-view 3D analysis using MEP indicates these columnar boundaries to be of low angle boundaries incapable of hosting chemical defects such as O, C impurities; this result will be published separately. Further, from a different TiN film grown and processed similar to the `AZ300T chip', microwave $T_{\text{c}} \simeq 5.1$ K is extracted from the quasi-particle model fit to the $Q_{\text{i}}$ vs $T$ data at multiple photon numbers. This matches the $T_{\text{c}} \simeq 5.1$ K extracted from direct current measurements shown in the inset of Fig.~\ref{fig:MatProps}(e). As shown by Crowley et al. \cite{crowley2023disentangling} using Ta resonators, reduced microwave $T_{\text{c}}$ is sometimes observed even for elemental superconductors. Twinned, columnar MBE TiN on Si with diffused metal-substrate interface also does not exhibit reduced microwave $T_{\text{c}}$ \cite{alexander2025power}. 

The columnar grains also result in surface undulations of the order 2 nm, the typical thickness of the tunnel barriers of Josephson junctions. Successful attempts to reduce the surface roughness using indium as surfactant have been made in the recent past in the context of binary and ternary transition metal nitrides \cite{lachowski2025nbn, lonergan2025indium}. TiN may also benefit from In surfactant but such a study is beyond the scope of this work.


{\bf {\em Acknowledgments:}} This material is based upon work supported by the Air Force Office of Scientific Research under award number FA9550-23-1-0688. 
Any opinions, findings, and conclusions or recommendations expressed in this material are those of the author(s) and do not necessarily reflect the views of the United States Air Force. This work was partially supported by an ONR Grant \# N00014-22-1-2633 monitored by Dr. Paul Maki. The authors acknowledge the use of facilities and instrumentation supported by NSF through the Cornell University Materials Research Science and Engineering Center DMR-1719875. This work was performed in part at the Cornell NanoScale Facility, an NNCI member supported by NSF Grant NNCI-2025233. The authors would like to acknowledge the fruitful discussions, regarding the resonator mask and participation ratio calculations, with Dr. Corey Rae McRae and Dr. Douglas Bennett of the National Institute of Standards and Technology, USA.

\clearpage
\bibliography{TiNresonator.bib}

\end{document}